\documentclass[aps,prd,twocolumn,10pt,superscriptaddress,nofootinbib,nobibnotes,longbibliography]{revtex4-1}

\usepackage{amssymb}
\usepackage{graphicx}
\usepackage{amsmath}
\usepackage{hyperref}
\usepackage{subfigure}
\usepackage{multirow}
\usepackage{setspace}
\usepackage{verbatim}
\usepackage{float}
\usepackage{color}
\usepackage[normalem]{ulem}
\usepackage[utf8]{inputenc}
\usepackage[table,xcdraw]{xcolor}
\usepackage{makecell}
\usepackage{url}
\usepackage{bm}
\usepackage{placeins}
\usepackage{slashed}

\newcommand{\md}{\mathrm{d}}

\begin{document}

%\title{Mass of Kerr-Newman black hole in the Bertotti-Robinson universe}
\title{Integrable mass of Kerr--Newman--Bertotti--Robinson black hole}

\author{Li Hu}
%\email{huli@nbu.edu.cn}
\affiliation{Institute of Fundamental Physics and Quantum Technology \& School of Physical Science and Technology, Ningbo University, Ningbo 315211, China}
\affiliation{Zhejiang Key Lab of Extreme Universe, Ningbo University, Ningbo 315211, China}

\author{Rong-Gen Cai}
%\email{caironggen@nbu.edu.cn}
\affiliation{Institute of Fundamental Physics and Quantum Technology \& School of Physical Science and Technology, Ningbo University, Ningbo 315211, China}
\affiliation{Zhejiang Key Lab of Extreme Universe, Ningbo University, Ningbo 315211, China}

\author{Shan-Wen Jiang}
%\email{2511690033@nbu.edu.cn}
\affiliation{Institute of Fundamental Physics and Quantum Technology \& School of Physical Science and Technology, Ningbo University, Ningbo 315211, China}
\affiliation{Zhejiang Key Lab of Extreme Universe, Ningbo University, Ningbo 315211, China}

\author{Shao-Jiang Wang}
\email{schwang@itp.ac.cn}
\affiliation{Institute of Theoretical Physics, Chinese Academy of Sciences (CAS), Beijing 100190, China}
\affiliation{Asia Pacific Center for Theoretical Physics (APCTP), Pohang 37673, Korea}

\begin{abstract}
The nontrivial asymptotic structure of rotating black holes in the Bertotti--Robinson universe complicates a direct determination of their conserved mass, leading previous studies to rely on a priori thermodynamic input. We address this long-standing problem using integrability methods, establish local analytic existence and uniqueness via the Cauchy--Kovalevskaja theorem, and obtain a closed expression via a rescaled recurrence whose termination is proved at all orders. The resulting Kerr--Newman--Bertotti--Robinson mass coincides with the expression previously derived using the canonical first law identity and reproduces the masses of both Kerr--Bertotti--Robinson subcases. With the mass thus fixed without invoking thermodynamic relations for the first time, we rigorously establish the first law of black-hole thermodynamics and the Smarr formula for a whole family of Kerr--Bertotti--Robinson black holes suitable for future applications.
\end{abstract}
\maketitle

\section{Introduction} 

Black holes are among the most fascinating predictions of general relativity, and recent studies continue to advance both theoretically and phenomenologically (e.g., Refs.~\cite {Ruffini:2024dwq, Hu:2025bbc, Lui:2026uai, Hu:2023oiu, Hu:2024hzu}). A consistent description of their physical properties requires a well-defined energy, yet assigning a conserved mass in spacetimes with nontrivial asymptotic structure remains an open problem.

Classical constructions such as the Komar integral and the Arnowitt-Deser-Misner (ADM) mass rely crucially on asymptotic flatness~\cite {Komar:1958wp, Arnowitt:1961zz}. More recent covariant phase space methods provide a framework not restricted to asymptotically flat geometries, allowing integrability to constrain the conserved mass~\cite {Iyer:1994ys, Barnich:2001jy, Barnich:2003xg}. For asymptotically anti-de Sitter (AdS) spacetimes, the Ashtekar--Das conformal method defines finite charges directly from the asymptotic Weyl tensor~\cite{Ashtekar:1999jx}. Applications to rotating AdS black holes~\cite{Gao:2023luj}, accelerating black holes~\cite{Appels:2016uha}, and Kerr-Newman-Melvin black holes~\cite{Astorino:2016hls} highlight the effectiveness of these modern approaches. Recently, rotating black holes in the Bertotti-Robinson (BR) universe provide a larger class of exact Einstein-Maxwell solutions, whose nontrivial asymptotic structure makes the determination of the conserved mass independent of thermodynamic input particularly challenging~\cite{Podolsky:2025tle, Ovcharenko:2026tos,Hu:2026slp}.

The BR universe has a conformally flat $\mathrm{AdS}_2\times S^2$ geometry supported by a homogeneous electromagnetic field~\cite{Bertotti:1959pf, Robinson:1959ev}, providing a distinct background from the Melvin magnetic universe~\cite{Melvin:1963qx}. The rotating black holes constructed in this background retain Petrov type D and admit a generalization to the four-parameter Kerr-Newman-Bertotti-Robinson (KN-BR) family~\cite{Podolsky:2025tle, Ovcharenko:2026tos}. Consequently, these solutions have recently attracted considerable attention and motivated numerous studies with various applications~\cite{Wan:2026lca, Astorino:2026kuv, Ovcharenko:2026pow, Zhang:2026myb, Ma:2026otg, DiPinto:2026rvp, Ma:2026ima, Xu:2026ags, Furugori:2026pdt, Hu:2026way, Roy:2026poj, Lyu:2026pqh, Hassanabadi:2026fzz, Li:2026gvm, Zhou:2026tkm, Chen:2026eqa, Wan:2026bkb, Zhou:2026fvj, Furugori:2026qbb, Ma:2026rro, Zhou:2026xrc, Wang:2026srw, Ma:2026gdn, Kutlu:2026zqu}.

%The conserved masses have already been obtained for the specially charged Kerr-BR$_s$, genuinely neutral Kerr-BR$_0$, and general KN-BR~\cite{Hu:2026slp, Kubiznak:2026uro, Zhang:2026sgj}, either by adopting the Christodoulou-Ruffini (CR) assumption or by imposing the condition that the first law holds. This means that previous studies cannot be said to have verified the first law, as the laws are automatically maintained or presupposed. 

The conserved masses of the specially charged Kerr-BR$_s$, genuinely neutral Kerr-BR$_0$, and general KN-BR black holes have previously been determined using thermodynamic input, including the Christodoulou--Ruffini (CR) assumption or the requirement that the first law hold~\cite{Hu:2026slp, Kubiznak:2026uro, Zhang:2026sgj}. However, the validity of these assumptions and the uniqueness of the resulting mass remain unknown. On the other hand, recovering thermodynamic laws within such constructions does not, by itself, constitute a verification independent of the thermodynamic input used to determine the mass.

In this Letter, we determine the conserved mass of the KN-BR black hole directly from the Barnich--Brandt integrability method, together with the Kerr-Newman (KN) normalization at the origin $B=0$. Then, an appropriate change of coordinates on the solution space reduces the integrability conditions to a polynomial master equation. Since the KN initial hypersurface is non-characteristic, the Cauchy--Kovalevskaja theorem guarantees a unique local analytic solution around $B=0$~\cite{Ebert2018}. To construct this solution explicitly, we derive a recurrence relation for a rescaled squared mass and obtain a closed-form expression. The resulting mass agrees with the KN-BR expression previously obtained from the canonical first law. This agreement not only independently validates the CR assumption for this black hole family but also, for the first time, enables us to verify the first law of black-hole thermodynamics and the Smarr formula using a mass determined without any thermodynamic input.

%The resulting mass agrees with the previously reported KN-BR mass and independently confirms the CR relation for this family. The mass is constructed without invoking the CR relation or a horizon first-law identity, after which the reconstructed generator allows an explicit check of the first law and the Smarr formula.

\section{KN-BR black hole}

The KN-BR solution is characterized by four parameters $(m, a, e, B)$, representing mass, rotation, electric charge, and external magnetic field, respectively. After normalizing the azimuthal coordinate to have the usual period $2\pi$, the metric takes the form~\cite{Ovcharenko:2026tos}
\begin{align}
    \md s^2&=\frac{1}{\Sigma^2}\bigg[-\frac{\beta}{\rho^2}\big(\md t-\frac{a\sin^2\theta}{P_0}\,\md\varphi\big)^2+\frac{\rho^2}{\beta}\md r^2+\frac{\rho^2}{P}\md\theta^2\nonumber\\
    &\qquad\qquad+\frac{P}{\rho^2}\sin^2\theta\,\big(a\md t-\frac{r^2+a^2}{P_0}\md\varphi\big)^2\bigg]\,,
\end{align}
where
\begin{align}
    \rho^2&=r^2+a^2\,\cos^2\theta\,,\qquad P=1+B^2\zeta^2\cos^2\theta\,,\nonumber\\
    \beta&=I\Delta\,,\qquad\Sigma^2=I-B^2\Delta\cos^2\theta\,,\nonumber\\
    I&=(1+k\,B\,r)^2+B^2r^2\,,\qquad P_0=1+B^2\zeta^2\,,\nonumber\\
    \Delta&=u\,a^2-2m\,r+(u-e^2/a^2)\,r^2\,,
\end{align}
and the auxiliary parameters obey
\begin{align}\label{eq:aux}
    u&=1+k^2\,,\quad k=\frac{e\,s-a\,m\,B}{a\,(1+e^2B^2)}\,,\nonumber\\
    \zeta^2&=m^2-u^2a^2+u\,e^2\,,\nonumber\\
    s^2&=1+(e^2-a^2)B^2-(m^2+e^2)a^2B^4\,.
\end{align}

After a constant gauge shift that ensures regularity along the rotation axis, $A_\varphi|_{\theta=0,\pi}=0$, the Maxwell potential can be described as~\cite{Zhang:2026sgj}
\begin{align}
A_\mu \md x^\mu&=\frac{a\,\md t}{\Sigma}\Big[-u\,B+\frac{r}{\rho^2}(m\,B\cos^2\theta-k)\Big]\nonumber\\&+\frac{\md \varphi}{P_0\Sigma}\Big[u\,a^2B+\frac{r}{\rho^2}\big(-m\,B(r^2+a^2)\cos^2\theta\nonumber\\&\quad+k\,a^2\sin^2\theta\big)-kr-\frac{1}{B}\Big]+\frac{\md \varphi}{P_0B}\frac{s+ameB^3}{1+e^2B^2}\,.
\end{align}
Together with the metric, the resulting electromagnetic field $F_{\mu\nu}=\partial_\mu A_\nu-\partial_\nu A_\mu$ exactly solves the Einstein-Maxwell equations.

For the allowed parameter range, we follow Ref.~\cite{Ovcharenko:2026tos} in choosing $s=\sqrt{s^2}>0$ and $\zeta=\sqrt{\zeta^2}>0$. The horizon radii are determined by the roots of $\Delta$,
\begin{align}
    r_\pm=\frac{m\pm\zeta}{u-e^2/a^2}\,.
\end{align}
Requiring both the inner and outer horizons to be positive implies $u-e^2/a^2>0$ and $m>\zeta$.

%We then proceed to determine the thermodynamic quantities associated with the outer event horizon. The locations of the horizons are given by roots of the function $\Delta$,
%\begin{align}
%    r_\pm=\frac{m\pm\zeta}{u-e^2/a^2}\,.
%\end{align}
%Given that both inner and outer horizons are required to be positive, one immediately has $u-e^2/a^2>0$ and $m>\zeta$. For the outer-horizon generator $\chi=\partial_t+\Omega_H\partial_\varphi$, the angular velocity, electrostatic potential, Hawking temperature, and entropy are~\cite{Bekenstein:1973ur,Bardeen:1973gs,Ovcharenko:2026tos}
%\begin{align}
%    \Omega_{H}&=-\left.\frac{g_{t\varphi}}{g_{\varphi\varphi}}\right|_{r=r_+}=\frac{aP_0}{r_+^2+a^2}\,,\nonumber\\
%    \Phi_H&=-A_\mu\chi^\mu|_{r=r_+}
%    =\frac{e\,r_+}{r_+^2+a^2}\,,\nonumber\\
%    T_H&=\frac{\kappa}{2\pi}=\frac{\zeta}{2\pi}\frac{(1+k\,B\,r_+)^2+B^2r_+^2}{r_+^2+a^2}\,,\nonumber\\
%    S&=\frac{\mathcal{A}}{4}=\frac{\pi}{P_0}\frac{r_+^2+a^2}{(1+k\,B\,r_+)^2+B^2r_+^2}\,,
%\end{align}
%where $\kappa$ and $\mathcal{A}$ represent the surface gravity and horizon area, respectively.

\section{Conserved charges and the mass equation}

To turn these geometric data into conserved quantities, consider Einstein gravity coupled to Maxwell theory,
\begin{align}
    I_\mathrm{EM}=\frac{1}{16\pi}\int\md^4x\sqrt{-g}\,(R-F_{\mu\nu}F^{\mu\nu})\,.
\end{align}
A vector ﬁeld $\xi$ and a Maxwell gauge parameter $\lambda$ deﬁne an exact symmetry when
\begin{align}
    \mathcal{L}_{\xi}g_{\mu\nu}&=0\,,\\
    \mathcal{L}_{\xi}A_{\mu}+\partial_{\mu}\lambda&=0\,.
\end{align}
Within covariant phase space, the Barnich-Brandt constructions associate a surface charge 2-form with such a symmetry~\cite{Barnich:2001jy, Barnich:2003xg}. Integrating it along a path through the space of solutions defines a finite charge
\begin{align}\label{eq:finitecharge}
    \mathcal{Q}_{(\xi,\lambda)}[g,A;\bar{g},\bar{A}]=\int_{S^2}\int_{(\bar{g},\bar{A})}^{(g,A)}k_{(\xi,\lambda)}[\delta g',\delta A';g',A']\,.
\end{align}
Here, $(\bar g,\bar A)$ denotes a reference solution, and $S^2$ is a surface which encloses the horizon for fixed time and radii. The surface charge associated with the generic symmetry parameter $(\xi,\lambda)$ is given by~\cite{Barnich:2001jy, Compere:2009dp},
\begin{align}
    k_{(\xi,\lambda)}[\delta g,\delta A;g,A]=k_{\xi}^{\rm grav}+k_{(\xi,\lambda)}^F\,,
\end{align}
whose two contributions are
\begin{subequations}\label{eq:surCharge}
\begin{align}
    k_{\xi}^{\rm grav}&=\frac{1}{8\pi}(\md^2x)_{\mu\nu}\Big\{\xi^\nu\nabla^\mu h-\xi^\nu\nabla_\rho h^{\mu\rho}+\xi_\rho\nabla^\nu h^{\mu\rho}\nonumber\\
    &\quad+\frac{1}{2}h\nabla^\nu\xi^\mu-h^{\rho\nu}\nabla_\rho\xi^\mu\Big\}\,,\\
    k_{(\xi,\lambda)}^F&=\frac{1}{4\pi}(\md^2x)_{\mu\nu}\bigg\{-F^{\mu\nu}\delta A_\rho\xi^\rho-2\xi^\mu F^{\nu\rho}\delta A_\rho\nonumber\\
    &\quad+\left[2h^{\mu\rho}F_\rho^{\;\nu}-\delta F^{\mu\nu}-\frac{1}{2}hF^{\mu\nu}\right]\left(A_\sigma\xi^\sigma+\lambda\right)\bigg\}\,.
\end{align}
\end{subequations}
The conventions we choose are
\begin{align}
    (\md^2x)_{\mu\nu}&=\frac{1}{4}\epsilon_{\mu\nu\rho\sigma}\,\md x^\rho\wedge\md x^\sigma\,,\qquad\epsilon_{tr\theta\varphi}=1\,,\nonumber\\
    h_{\mu\nu}&=\delta g_{\mu\nu}\,,\qquad h^{\mu\nu}=g^{\mu\rho}g^{\nu\sigma}h_{\rho\sigma}\,,\qquad h=g^{\mu\nu}h_{\mu\nu}\,,\nonumber\\
    \delta F_{\mu\nu}&=\partial_\mu\delta A_\nu-\partial_\nu\delta A_\mu\,,\qquad\delta F^{\mu\nu}=g^{\mu\rho}g^{\nu\sigma}\delta F_{\rho\sigma}\,.\nonumber
\end{align}

For fixed symmetry parameters, the path independence of Eq.~\eqref{eq:finitecharge} requires the following integrability condition~\cite{Compere:2006my}
\begin{align}
    \int_{S^2}&\big(\delta_1 k_{(\xi,\lambda)}[\delta_2 g,\delta_2 A;g,A]\nonumber\\
    &-\delta_2 k_{(\xi,\lambda)}[\delta_1 g,\delta_1 A;g,A]\big)=0\,.
\end{align}
%For the axial symmetry $(-\partial_\varphi,0)$ and the constant gauge symmetry $(0,-1)$, the complete surface integrals give conserved angular momentum and electric charge
According to this formalism, the angular momentum and electric charge associated with the axial symmetry $(-\partial_\varphi,0)$ and the constant gauge symmetry $(\xi,\lambda)=(0,-1)$ can be obtained straightforwardly
\begin{align}
    J=\mathcal{Q}_{(-\partial_\varphi,0)}=\frac{ma}{P_0^2}\,,\qquad Q&=\mathcal{Q}_{(0,-1)}=\frac{e}{P_0}\,.
\end{align}
Unfortunately, when dealing with conserved mass, the charge corresponding to the coordinate time translation symmetry cannot be guaranteed to be integrable since the canonical generator is not known a priori. Here, we use a slash in the definition of the infinitesimal conserved charge $\omega$ to emphasize this failure to be a total differential form,
\begin{align}\label{eq:slash_mass}
\omega=\slashed{\delta}\mathcal{Q}_{(\partial_t,0)}=c_m\delta m+c_a\delta a+c_e\delta e+c_B\delta B\,.
\end{align}

Given that we have three independent generators, a natural approach is to consider a linear combination to characterize the conserved energy, as in the previous construction~\cite{Astorino:2016hls, Hu:2026slp},
\begin{align}\label{eq:squaredvariation}
\delta(M^2)=\lambda\,\omega+\eta\,\delta J+\tau\,\delta(Q^2)\,,
\end{align}
where $\delta(Q^2)=2Q\delta Q$ spans the same variation direction as $\delta Q$ while keeping the coefficients rational. The three multipliers are functions of the solution parameters, from which we reconstruct the canonical generator.

Before substituting the explicit variational coefficients, we introduce an auxiliary parameter $k$ in place of the electric parameter $e$ to eliminate the radicals involving $s$ and simplify the subsequent calculations. The auxiliary relation takes the following form,
\begin{align}\label{eq:constraint}
Z=1-ua^2B^2\,,\qquad
p\equiv e^2=\frac{a^2(k+mB)^2}{Z}\,.
\end{align}
This is a local coordinate change on the four-parameter solution space and imposes no additional physical condition. After that, the transformed coefficients are
\begin{equation}
\begin{aligned}
\widetilde c_m&=c_m+\frac{ec_e}{2p}\,\partial_m p=\frac{\Xi_m}{D^2}\,,\\
\widetilde c_a&=c_a+\frac{ec_e}{2p}\,\partial_a p=\frac{\Xi_a}{D^2}\,,\\
\widetilde c_k&=\frac{ec_e}{2p}\,\partial_k p=\frac{\Xi_k}{D^2}\,,\\
\widetilde c_B&=c_B+\frac{ec_e}{2p}\,\partial_B p=\frac{\Xi_B}{D^2}\,,
\label{eq:pullback}
\end{aligned}
\end{equation}
where the exact forms of the four numerators $\Xi_i$ are shown in the End Matter~\ref{app:coefficients}, and the common denominator is built from
\begin{align}
D=ZP_0=1+(m^2-2ua^2)B^2+2kuma^2B^3+u^3a^4B^4\,.
\end{align}

Expanding Eq.~(\ref{eq:squaredvariation}) now yields the following four equations
\begin{subequations}\label{eq:int_condi}
\begin{align}
    \partial_m M^2&=\lambda\,\tilde{c}_m+\eta\,\partial_m J+\tau\,\partial_m Q^2\,,\label{eq:int_condi_m}\\
    \partial_a M^2&=\lambda\,\tilde{c}_a+\eta\,\partial_a J+\tau\,\partial_a Q^2\,,\label{eq:int_condi_a}\\
    \partial_k M^2&=\lambda\,\tilde{c}_k+\eta\,\partial_k J+\tau\,\partial_k Q^2\,,\label{eq:int_condi_k}\\
    \partial_B M^2&=\lambda\,\tilde{c}_B+\eta\,\partial_B J+\tau\,\partial_B Q^2\,.\label{eq:int_condi_B}
\end{align}
\end{subequations}
Now that there are four equations and four unknowns, $(M^2,\lambda,\eta,\tau)$, this implies that, in principle, the conserved mass can be uniquely determined solely from the integrability condition. Eliminating the three multipliers $\lambda$, $\eta$, and $\tau$ leaves a differential constraint on the unknown mass $M^2$,
\begin{align}\label{eq:master}
\quad V(M^2)=0\,,\quad V=A_m\partial_m +A_a\partial_a 
 +A_k\partial_k +A_B\partial_B\,.
\end{align}
The coefficients $A_m$, $A_a$, $A_k$, and $A_B$ are polynomials in $B$ of degrees $4$, $4$, $4$, and $5$, respectively, with their explicit expressions given in the End Matter~\ref{app:coefficients}. Note that here we fix the normalization by setting $A_B|_{B\to 0}=m$, which amounts to an overall rescaling of the master equation. For $m>0$, this condition makes $B=0$ a non-characteristic hypersurface, allowing the conserved mass to be determined as an initial-value problem.

\section{Recursive construction of the mass}

%We now proceed to focus on the existence and uniqueness of solutions to Eq.~(\ref{eq:master}), as well as how to construct such a solution.

Considering that the KN limit supplies the analytic initial function $M^2(m,a,k,0)=m^2$, and the coefficients of $V$ are all polynomials. Then, according to the Cauchy–Kovalevskaja theorem~\cite{Ebert2018}, there is a unique real analytic solution defined in a neighborhood of $B=0$, whose expansion is
\begin{align}
    M^2(m,a,k,B)=m^2+\sum_{j\ge1}f_j(m,a,k)B^j\,.
\end{align}
However, this qualitative result, by itself, does not yield a useful closed-form expression. In fact, our calculations reveal that the direct expansion of $M^2$ is non-terminating, as confirmed by examining the coefficients at each order of the exact solution obtained below. We therefore turn to the rescaled function
\begin{align}
    W_n=D^nM^2=m^2+\sum_{\ell\ge1}w^{(n)}_\ell B^{\ell}\,.
\end{align}
The integer $n\ge0$ is initially undetermined. Since $D|_{B\rightarrow0}=1$, this change of unknown is invertible near $B=0$ for every fixed $n$. Consequently, it retains every analytic mass candidate.

Direct differentiation of $D$ gives
$V(D)=\Gamma D$, where the polynomial $\Gamma$, listed in the End Matter~\ref{app:coefficients}, has degree $4$ in $B$ and vanishes at $B=0$. The master equation is then converted into 
\begin{align}
    V(W_n)-n\Gamma W_n=0\,,
\end{align}
which can be decomposed based on the order of $B$. To display its coefficient equations, define
\begin{equation}
\begin{aligned}
A_\psi&=\sum_{j=0}^4A_{\psi,j}B^j\quad(\psi=m,a,k)\,,\\
A_B&=m+\sum_{j=1}^5b_jB^j\,,\qquad
\Gamma=\sum_{j=1}^4\gamma_jB^j\,.
\end{aligned}
\label{eq:coefficientdefs}
\end{equation}
By collecting the terms order by order in the master equation, we get
\begin{subequations}
\begin{align}
w_1^{(n)}&=2kuma^2\,,\\
w_2^{(n)}&=(n-3)m^4+(4+5k^2-2nu)m^2a^2+uk^4a^4\,,\\
w_{\ell}^{(n)}&=-\frac{1}{\ell m}\sum_{j=0}^{\min(4,\ell-1)}\bigg[
 A_{m,j}\partial_m w_{\ell-j-1}^{(n)}+A_{a,j}\partial_a w_{\ell-j-1}^{(n)}\nonumber\\
&+A_{k,j}\partial_k w_{\ell-j-1}^{(n)}+\big((\ell-j-1)b_{j+1}-n\gamma_j\big)w_{\ell-j-1}^{(n)}\bigg]\,.\label{eq:Wrecurrence}
\end{align}
\end{subequations}
For $\ell\ge5$, the degree bounds in Eq.~\eqref{eq:coefficientdefs} restrict each subsequent step to the five preceding coefficient functions $w_{\ell-1}^{(n)},\ldots,w_{\ell-5}^{(n)}$.

Keeping $n$ symbolic through the ninth step reveals
\begin{align}\label{eq:w9structure}
w_9^{(n)}=(n-3)\,\widehat w_9^{(n)}(m,a,k)\,.
\end{align}
The remaining polynomial factor is recorded in the End Matter~\ref{app:coefficients}. Although this factorization suggests the rescaling $W_3=D^3M^2$, one zero coefficient alone would not prove termination. Continuing the recurrence at $n=3$ yields five exact identities,
\begin{align}\label{eq:fivezeros}
w_9^{(3)}=w_{10}^{(3)}=w_{11}^{(3)}=w_{12}^{(3)}=w_{13}^{(3)}=0\,.
\end{align}
These coefficient functions and their parameter derivatives all vanish. At $\ell=14$, Eq.~\eqref{eq:Wrecurrence} involves only $w_{13}^{(3)},\ldots,w_9^{(3)}$, so $w_{14}^{(3)}=0$. Each following step again acts on five zero functions. Induction therefore proves that the entire tail vanishes, establishing an exact polynomial rather than a finite-order approximation.

The polynomial termination established above also highlights the importance of replacing $e$ with $k$ through Eq.~\eqref{eq:constraint}. In the original variables $(m,a,e,B)$, the coefficient functions contain $s=\sqrt{s^2}$. Although $s^2$ is polynomial in $B$, its square root generally has a nonterminating Taylor expansion, obscuring the finite-polynomial structure underlying our construction. The change of variables rationalizes these expressions and enables the closed-form reconstruction through the terminating recurrence.

The nonzero coefficients assemble into $W_3=ZS_6$, with
\begin{align}\label{eq:S6}
S_6&=m^2+2kuma^2B+(uk^4a^4-m^2a^2)B^2\nonumber\\
&+2kma^2(m^2-2ua^2)B^3\nonumber\\
&+m^2a^2[m^2+u(4k^2-1)a^2]B^4\nonumber\\
&+2kuma^4(m^2+u^2a^2)B^5+u^3m^2a^6B^6\,.
\end{align}
After appropriate simplification, the resulting mass, $M^2=S_6/(Z^2P_0^3)$, takes the compact form
\begin{align}\label{eq:mass}
\boxed{M^2=\frac{m^2+2kuma^2B+B^2(ue^4+m^2e^2+um^2a^2)}{P_0^3}\,.}
\end{align}
Here, $k$ and $e$ remain related by Eq.~\eqref{eq:aux}, and $M|_{B\rightarrow 0}=m>0$ selects the positive root. 

This result agrees with the KN-BR mass reported in Ref.~\cite{Zhang:2026sgj} and reproduces the Kerr-BR$_0$ and Kerr-BR$_s$ masses under the corresponding parameter restrictions~\cite{Kubiznak:2026uro, Hu:2026slp}. The agreement independently confirms the validity of the CR assumption for this family. Whereas previous derivations relied on thermodynamic input, the present construction determines the mass solely from the integrability condition, allowing us in turn to verify the first law and the Smarr formula independently.

\section{Verification of thermodynamic laws}

To maintain formal consistency with the first law of black-hole thermodynamics, the projections of the conserved energy along the three Killing vectors can be reformulated as~\cite{Hu:2026slp}
\begin{align}\label{eq:generatorlaw}
\delta M=\alpha(\omega-\Omega_{\rm int}\delta J-\Phi_{\rm int}\delta Q)\,.
\end{align}
Comparison with Eq.~\eqref{eq:squaredvariation} identifies $\alpha=\lambda/(2M)$, $\Omega_{\rm int}=-\eta/\lambda$, and $\Phi_{\rm int}=-2Q\tau/\lambda$. Solving the component equations with the conserved mass gives
\begin{align}\label{eq:generatorcoefficients}
\alpha&=\frac{ZL}{MD^2}\,,\nonumber\\
\Omega_\mathrm{int}&=\frac{aB(k+mB)P_0[-1+kmB+u^2a^2B^2]}{L}\,,\nonumber\\
\Phi_{\rm int}&=-\frac{eB(k+mB)(m+kua^2B)}{L}\,,
\end{align}
where
\begin{align}\label{eq:L}
L=m&+k^3a^2B-2ma^2B^2+km^2a^2B^3+mu^2a^4B^4\,.
\end{align}

%We then proceed to determine the thermodynamic quantities associated with the outer event horizon. 

For the outer-horizon generator $\chi=\partial_t+\Omega_H\partial_\varphi$, the angular velocity, electrostatic potential, Hawking temperature, and entropy are~\cite{Bekenstein:1973ur,Bardeen:1973gs,Ovcharenko:2026tos}
\begin{align}
    \Omega_{H}&=-\left.\frac{g_{t\varphi}}{g_{\varphi\varphi}}\right|_{r=r_+}=\frac{aP_0}{r_+^2+a^2}\,,\nonumber\\
    \Phi_H&=-A_\mu\chi^\mu|_{r=r_+}
    =\frac{e\,r_+}{r_+^2+a^2}\,,\nonumber\\
    T_H&=\frac{\kappa}{2\pi}=\frac{\zeta}{2\pi}\frac{(1+k\,B\,r_+)^2+B^2r_+^2}{r_+^2+a^2}\,,\nonumber\\
    S&=\frac{\mathcal{A}}{4}=\frac{\pi}{P_0}\frac{r_+^2+a^2}{(1+k\,B\,r_+)^2+B^2r_+^2}\,,
\end{align}
where $\kappa$ and $\mathcal{A}$ represent the surface gravity and horizon area, respectively.

With the thermodynamic potentials redefined as follows,
\begin{align}\label{eq:canonicalpotentials}
T&=\alpha T_H\,,\qquad \Omega=\alpha(\Omega_H-\Omega_{\rm int})\,,\nonumber\\
\Phi&=\alpha(\Phi_H-\Phi_{\rm int})\,,
\end{align}
one readily verifies both the standard first law of black-hole thermodynamics and the Smarr formula,
\begin{align}
    \delta M&=T\delta S+\Omega\delta J+\Phi\delta Q\,,\nonumber\\
    M&=2TS+2\Omega J+\Phi Q\,.
\end{align}

\section{Conclusions and discussions}

We have determined the conserved mass of the KN-BR black hole directly from the Barnich-Brandt constructions, using the independently computed angular momentum and electric charge together with the KN normalization at $B=0$. Neither the CR relation nor the first law is imposed in this construction. A change of coordinates on the solution space converts the integrability condition into a polynomial master equation. Its non-characteristic KN initial hypersurface allows the Cauchy--Kovalevskaja theorem to establish local analytic existence and uniqueness. The rescaled recurrence then constructs this solution explicitly. Because each new coefficient depends on at most five preceding coefficient functions, five consecutive identically vanishing coefficients establish termination of the series and yield a closed expression for the mass.

The unique resulting mass agrees with the earlier KN-BR expression and reproduces the Kerr-BR$_s$ and Kerr-BR$_0$ masses under the corresponding parameter restrictions. These agreements provide independent checks of the earlier mass assignments and of the CR relation for this family. Reconstructing the mass generator and combining it with the thermodynamic quantities defined at the horizon subsequently verifies the first law and the Smarr formula without any thermodynamic input. Within this scope, the construction fixes the mass ambiguity and separates its determination from the thermodynamic relations that it ultimately satisfies.

\vspace{0.8\baselineskip}
\begin{acknowledgments}
This work is supported by the National Key Research and Development Program of China Grants (No. 2021YFC2203004 and No. 2021YFA0718304), the National Natural Science Foundation of China Grants (No. 12422502, No. 12547110, No. 12588101, No. 12235019, and No. 12447101).
\end{acknowledgments}

\bibliography{biblatex}

\clearpage
\onecolumngrid
\appendix
\numberwithin{equation}{section}
\renewcommand{\theequation}{\thesection\arabic{equation}}
\section*{End Matter: Exact polynomial expressions}
\label{app:coefficients}

The numerators of the transformed coefficients in Eq.~\eqref{eq:pullback} are
\begin{subequations}\label{app:Xi}
\begin{align}
\Xi_m&=Z\big[1-(2m^2+ua^2)B^2-2kuma^2B^3\big]\,,\\
\Xi_a&=uaB\Big\{k+3mB-5km^2B^2-m\big[3m^2+2u(3u-1)a^2\big]B^3
-kua^2(2m^2+u^2a^2)B^4+u^3ma^4B^5\Big\}\,,\\
\Xi_k&=a^2B\Big\{2u-1+3kmB-\big[(7u-5)m^2+2u^2a^2\big]B^2
-km\big[3m^2+2u(4u-1)a^2\big]B^3\nonumber\\
&\hspace{2.7em}+\big[2um^2a^2+u^3(2-u)a^4\big]B^4+3ku^3ma^4B^5\Big\}\,,\\
\Xi_B&=kua^2+m(4ua^2-3m^2)B-9kum^2a^2B^2-u^2(7u-1)ma^4B^3
+ku^2a^4(2m^2-u^2a^2)B^4+2u^4ma^6B^5\,.
\end{align}
\end{subequations}
The polynomial coefficients of the master equation are
\begin{subequations}\label{app:mastercoefficients}
\begin{align}
A_m&=-kuma^2+m^2\big[3m^2-(1+2u)a^2\big]B
+kma^2\big[(1+9u)m^2-u^2a^2\big]B^2\nonumber\\
&+m^2a^4u(1-7u+9u^2)B^3-kma^4u\big[m^2+u^2(1-3u)a^2\big]B^4\,,\label{app:Am}\\
A_a&=kua^3+ma\big[m^2+(1-2u)a^2\big]B
+ka^3\big[(3u-1)m^2-3u^2a^2\big]B^2\nonumber\\
&+ma^5u(-1-u+3u^2)B^3+ka^5u\big[m^2+u^2(1+u)a^2\big]B^4\,,\label{app:Aa}\\
A_k&=-m^2+u(1-u)a^2+km(m^2-2ua^2)B+m^2a^2u(3u-1)B^2\nonumber\\
&+kma^4u^2(3u-1)B^3-a^4u^2\big[m^2+u^2(1-u)a^2\big]B^4\,,\label{app:Ak}\\
A_B&=m+k(u-1)a^2B-m\big[2m^2+(2-u)a^2\big]B^2
-ka^2\big[(6u-2)m^2-u^2a^2\big]B^3\nonumber\\
&+ma^2\big[m^2+(-u+7u^2-6u^3)a^2\big]B^4
+ka^4u\big[m^2+u^2(1-2u)a^2\big]B^5\,.\label{app:AB}
\end{align}
\end{subequations}
The polynomial in $V(D)=\Gamma D$ is
\begin{align}\label{app:Gamma}
\Gamma=2B(m+kua^2B)\Big[m^2-2ua^2+kma^2(1+2u)B
+a^2\big\{m^2+u^2(1+u)a^2\big\}B^2\Big]\,.
\end{align}
Finally, the remaining factor in Eq.~\eqref{eq:w9structure} is
\begin{align}\label{app:w9}
\widehat w_9^{(n)}=\frac{km^9a^2}{12}
\big[p_0-4xp_1+4x^2p_2-8x^3p_3+4x^4p_4\big]\,,
\qquad x=\frac{ua^2}{m^2}\,,
\end{align}
where the dimensionless polynomials $p_i$, distinct from the metric function $P_0$, are
\begin{subequations}\label{app:p9}
\begin{align}
p_0&=(n-4)(n-5)(5un-26-30k^2)\,,\\
p_1&=(n-4)\big[8un^2-(73+84k^2)n+159+220k^2\big]\,,\\
p_2&=18un^3-(207+218k^2-28k^4)n^2+(768+909k^2-252k^4)n-924-1300k^2+560k^4\,,\\
p_3&=8un^3-(74+72k^2-30k^4)n^2+(217+235k^2-255k^4)n-205-288k^2+540k^4\,,\\
p_4&=4un^3-(24+28k^2-24k^4)n^2+(47+77k^2-177k^4+9k^6)n-30-80k^2+330k^4-36k^6\,.
\end{align}
\end{subequations}

\end{document}